\def   \ni {\noindent}

\def   \ssk {\vskip  5truept}

\def   \bsk {\vskip 15truept}
 
\def   \newpage {\vfill\eject}
\def   \newline {\hfil\break}

\newcommand{\nel}{n_{\mathrm{e}}}
\newcommand{\thom}{\sigma_{\mathrm{T}}}

\newcommand{\ktr}{kT_{\mathrm{r}}}
\newcommand{\tauh}{\tau_{\mathrm{H}}}
\newcommand{\tor}{\tau_{\mathrm{R}}}

\documentstyle[epsfig]{article}
\begin{document}

\hsize 5truein
\vsize 8truein
\font\abstract=cmr8
\font\keywords=cmr8
\font\caption=cmr8
\font\references=cmr8
\font\text=cmr10
\font\affiliation=cmssi10
\font\author=cmss10
\font\mc=cmss8
\font\title=cmssbx10 scaled\magstep2
\font\alcit=cmti7 scaled\magstephalf
\font\alcin=cmr6 
\font\ita=cmti8
\font\mma=cmr8
\def\ref{\par\noindent\hangindent 15pt}
\null


\title{\ni IC EMISSION ABOVE A DISK}

\bsk \bsk
\author{\ni J.~Malzac, E.~Jourdain
}                                                       
\bsk
\affiliation{C.E.S.R. (CNRS/UPS), 9, av. du Colonel Roche, B.P. 4346, 31028 Toulouse Cedex 4, France 
}                                                
\bsk
\baselineskip = 12pt

\abstract{ABSTRACT \ni

We developed a non-linear Monte Carlo code based on the LP
 method proposed by Stern et al (1995).
We use it to simulate an extended non-thermal X/$\gamma$ ray
source coupled with a cold accretion disk.
The source radiates by inverse Compton process.
 We show that very low optical depths 
($\tau\sim5.10^{-4}$) are required for the 2nd Compton order to be negligible.
Such optical depths can be reached only if the source is very close to the disk (corona-like geometries). And even in this configuration, 2nd Compton order is negligible only for the hardest observed spectral slopes.  
We also account for Compton reflection on the disk. 
This kind of coupling leads to a strongly angle
 dependent spectrum with a reflection component much
 stronger than in standard isotropic models, leading to unobserved reflection dominated spectra at low inclination.}
\bsk
\baselineskip = 12pt
\keywords{\ni KEYWORDS: GBHCs, Seyfert Galaxies, radiative transfer, X-rays, $\gamma$-rays
}

\bsk
\baselineskip = 12pt


\text{\ni 1. INTRODUCTION
\ssk
\ni   
  
The high energy spectra of Seyfert galaxies and black hole binaries in their soft hard state are very similar. Their basic characteristics are an underlying powerlaw with photon index $\alpha\sim1.6-2.$, a Compton reflection component with an Fe $K\alpha$ line and a high energy cutoff above $\sim200$ \,keV.
The powerlaw is generally though to originate through Comptonisation of soft photons by a population of thermal hot electrons. Here we will focus on non-thermal models following Henri \& Petrucci (1997, HP97) who proposed a model where the hard X-rays are emitted by  a non-thermal optically thin point source. This model reproduces the basic features of the high energy spectra, but requires high inclination angle for all systems unless one assumes that a fair part of the disk is highly ionised (Malzac et al. 1998).
 However considering a point source, as done in those works, is unrealistic (HP97) As the source was assumed to be optically thin the second Compton scattering order (CSO) was neglected and the calculations performed in the Thomson limit. Here we developed a Non Linear Monte Carlo Code (inspired from Stern et al., 1995) to investigate the effects of source extension taking properly into account the radiative transfer and the Klein Nishina effects.
                     
\begin{figure}
\centerline{
\psfig{file=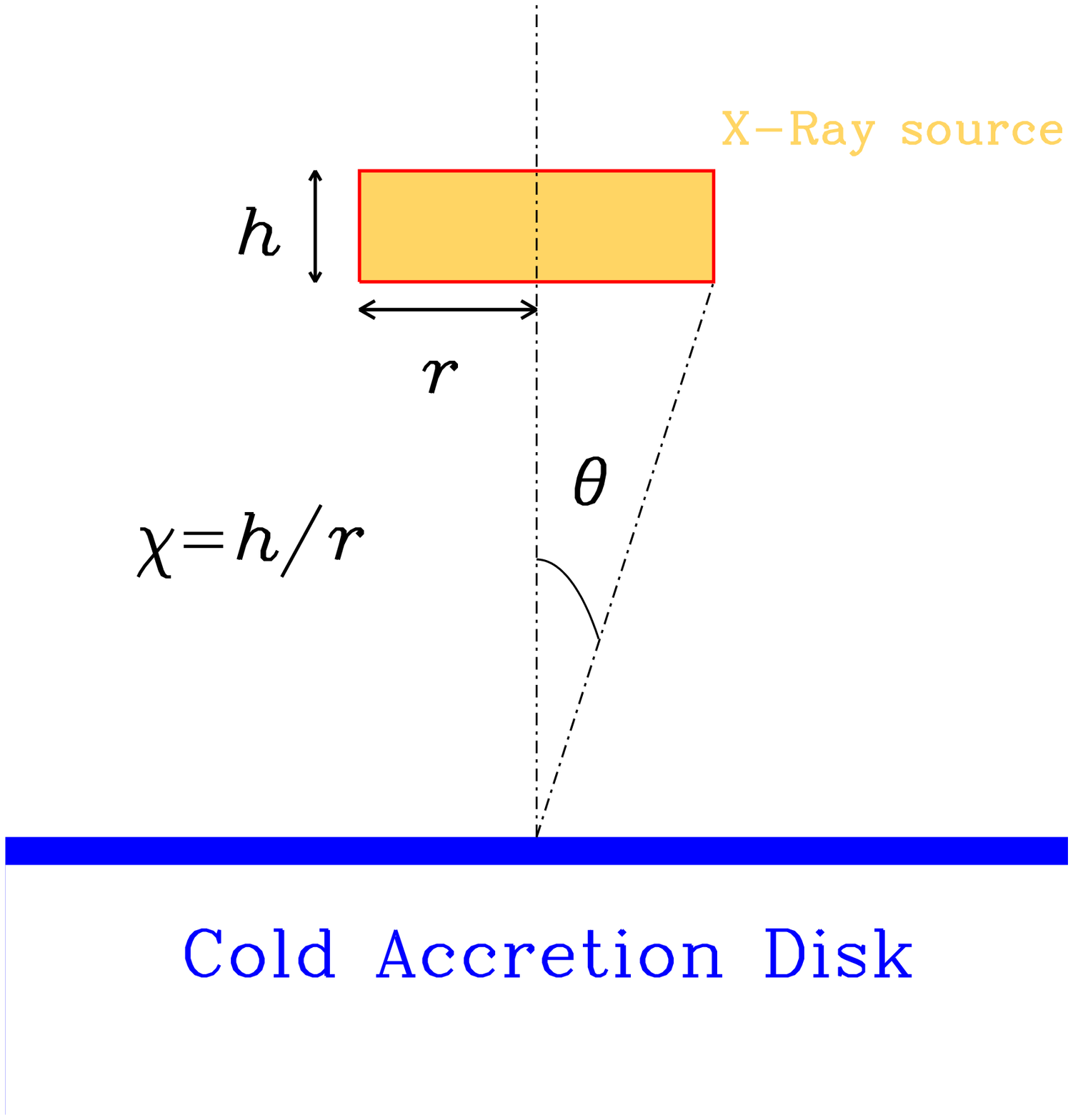, width=4.5cm}
\hspace{2cm}\psfig{file=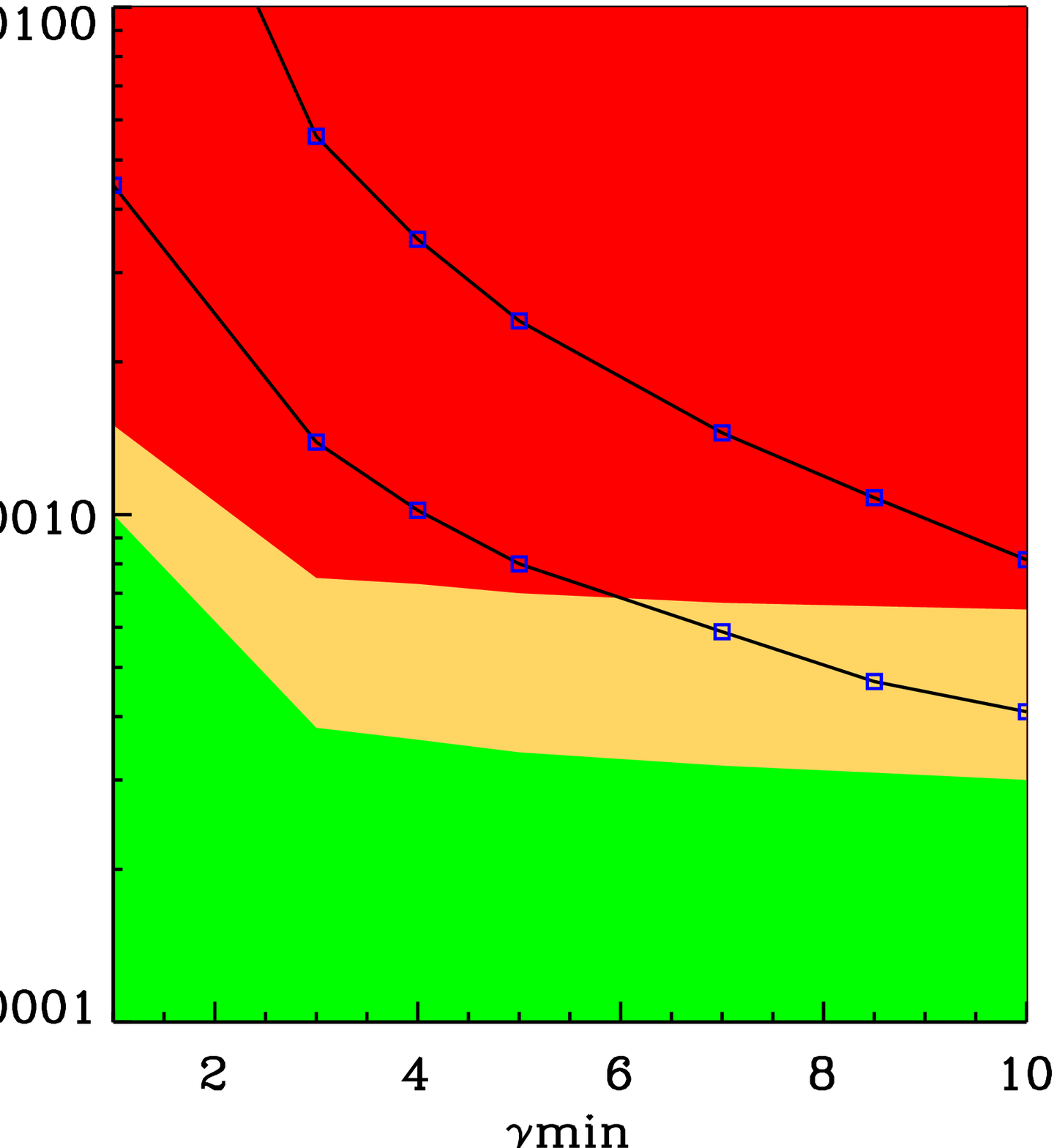,width=4.5cm}
}
\centerline{
\caption{FIGURE 1.}
\hspace{5cm}\caption{FIGURE 2.}
}
\end{figure}
\newpage
\bsk
\ni 2. MODEL ASSUMPTIONS 
\ssk
\ni 
 The X-ray source is modeled by a cylinder at some height above a disk. This geometry could be physically realized by a strong shock terminating an aborted jet (Henri \& Petrucci 1997). The source radiates by inverse Compton process illuminating the disk. The seed photons are provided by the thermal disk emission at fixed blackbody temperature $\ktr$. This disk is represented by an infinite slab which radiates only the reprocessed energy from the hot source without internal dissipation. In addition a fraction of the high energy photons intercepting the disc is Compton reflected before being absorbed leading to the formation of the so called reflection component. 
 The relevant geometric parameters characterizing the source are $\chi$ and $\theta$ (see Fig.~1).
The cylinder contains relativistic electrons or pairs with a fixed energy distribution:
\begin{eqnarray}
f(\gamma)=A\:\gamma^{-s}e^{-\left(\frac{\gamma}{\gamma_{0}}\right)^{\beta}}
           \,\,\,\,\,for\, \gamma>\gamma_{min}
\end{eqnarray}
These leptons cool on the soft photon field. We assume that some acceleration process is active in the X-ray source to keep the above distribution fixed. The equilibrium cylinder optical depth depends only on the geometry and electron distribution and remains to be determined self-consistently. The results are scale and luminosity independent since we neglected photon-photon pair production and pair annihilation.

The distribution parameters are constrained by the observations: $2<s<3$, $\gamma_{0}<200$, $\ktr>1 eV$. $\gamma_{min}<10$ otherwise the lack of low energy electrons would cause an unobserved  break above 1 keV in the photon spectra. The most drastic contraint arises from the requirement of the second and higher CSOs to be negligible. Indeed, due to the high mean electron energy, the combination of CSOs leads to a photon spectrum extending well above $1 MeV$. Thus the model requires very small optical depths typically $\tau\sim 5.10^{-4}$ (green and yellow region of Fig. 2). The relative strengh of the 2nd CSO depends also on $s$, $\gamma_{min}$ and the viewing angle (Fig. 5,6).

\begin{figure}
\centerline{
\psfig{file=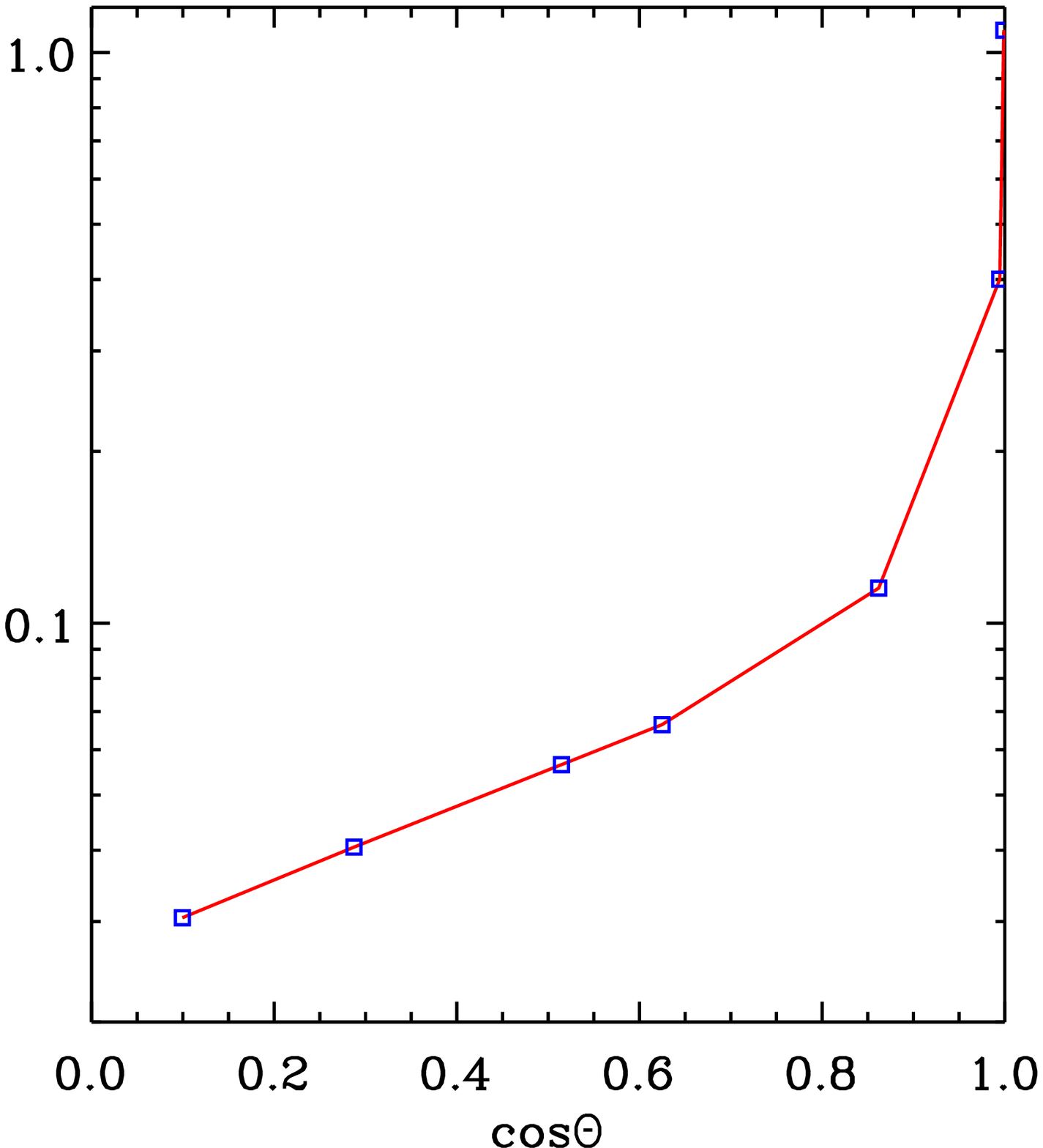,width=4cm}
\hspace{3cm}\psfig{file=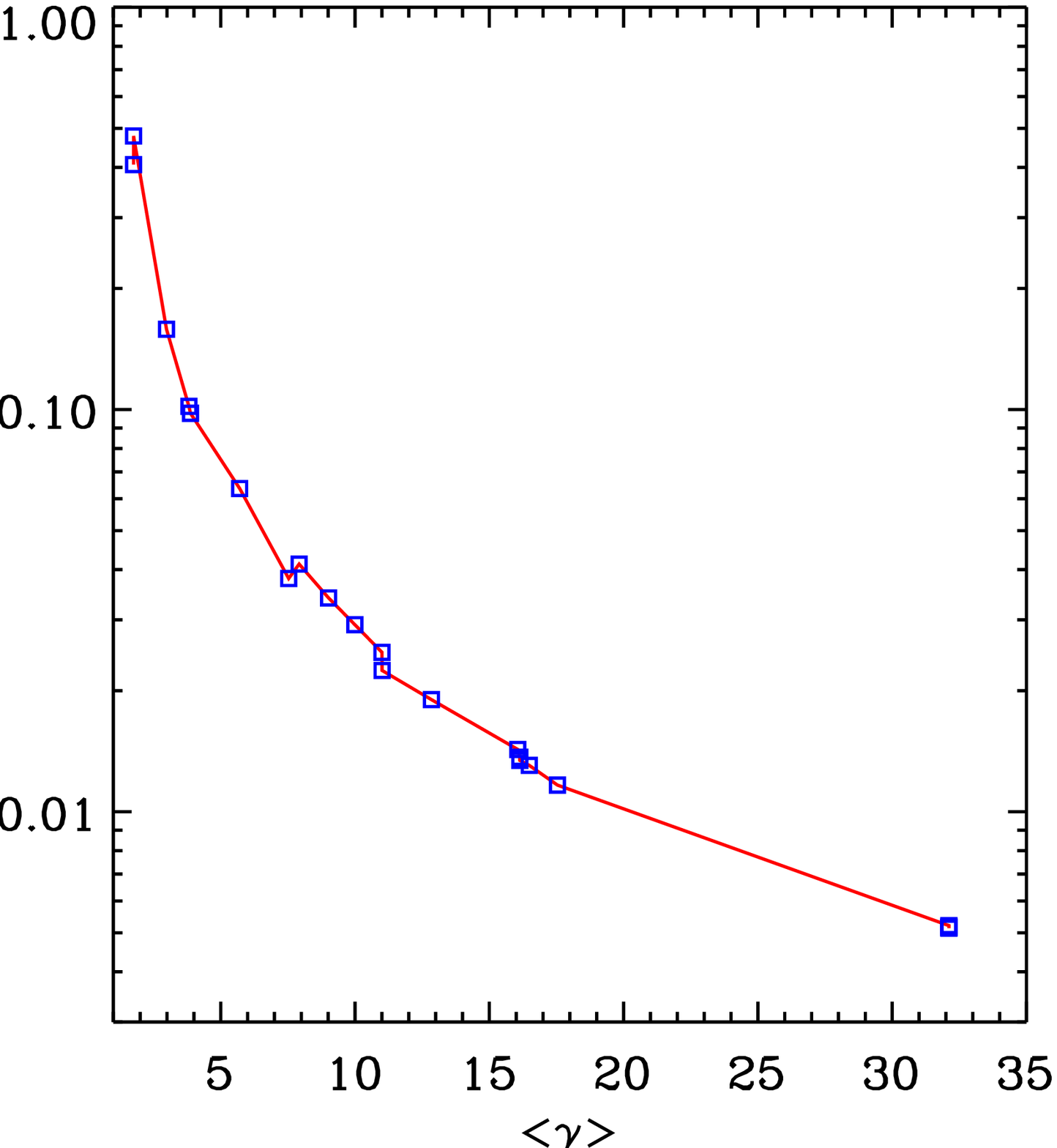,width=4cm}
}
\centerline{
\caption{FIGURE 3.}
\hspace{5cm}\caption{FIGURE 4. }
}
\end{figure}

\newpage 
3. RESULTS 
\ni
\ssk
\ni

 Our NLMC code computes both the optical depth  and the resulting photon spectra. This code has been tested successfully against a linear code (Podzniakov et al, 1983). In this version of the code, the power supplied to the electrons as well as their distribution are fixed. At each time step the power is supplied in the form of an injection of electrons. In balance, the number of electrons is decreased according to the Compton losses. So that, the optical depth 'evolves' until it reaches an equilibrium. We performed over 100 simulations covering a wide range of parameter values. The optical depth can be defined along the height of the cylinder or along its radius:
 \begin{eqnarray}
\tauh=\frac{\nel}{V}\thom H=K\left(1+\chi\right)
\hspace{3cm}\tor=\frac{\nel}{V}\thom R =K\left(1+1/\chi\right)
\end{eqnarray}
where $K$ is a function of $\theta$, $ \chi$, $ \gamma_{min}$, $ \gamma_{0}$, $ s$, $ \beta$ and $ kTr$. $K$ is the lowest optical depth that can be achieved in any direction.

Using our NLMC simulations we studied the dependence of $K$ on these parameters. While $\chi$ has a slight influence on $K$, increasing $K$ of at most a factor of 2 between $\chi=10^{-3}$ and $\chi=10^{3}$, $K$ is very sensible to the solid angle sustended by the cylinder as seen from the source. Fig. 3 shows $K$ as a function of $cos\theta$ $K$ for $\chi=1$, $\gamma_{0}=50$, $\gamma_{min}=2$, $\beta=2$, $s=3$, $ktr=50 $eV. $K$ decreases strongly with $\theta$, minimum values are obtained for $\theta\sim\pi/2$.
\begin{figure}[h]
\centerline{\psfig{file=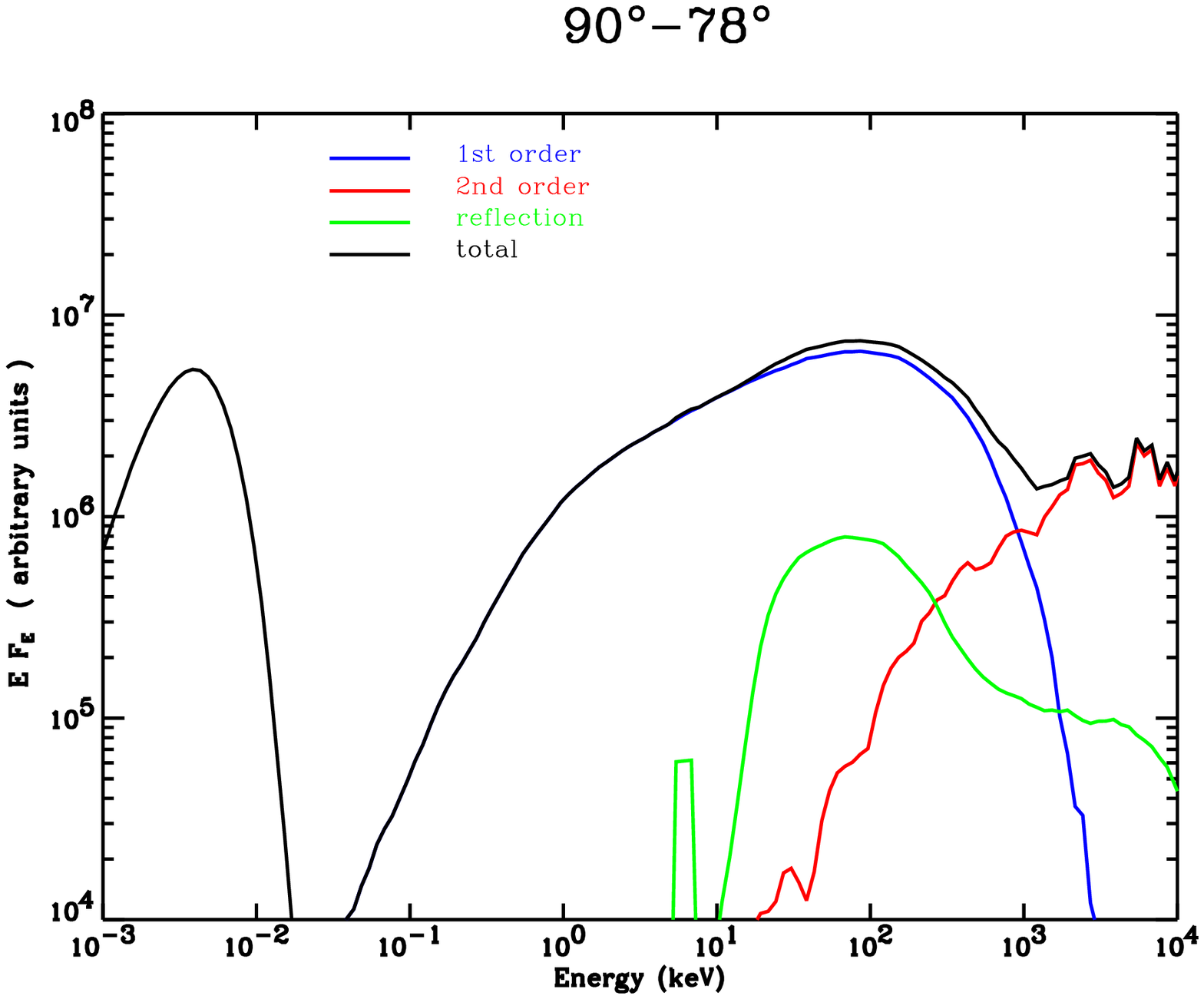,width=7cm}\psfig{file=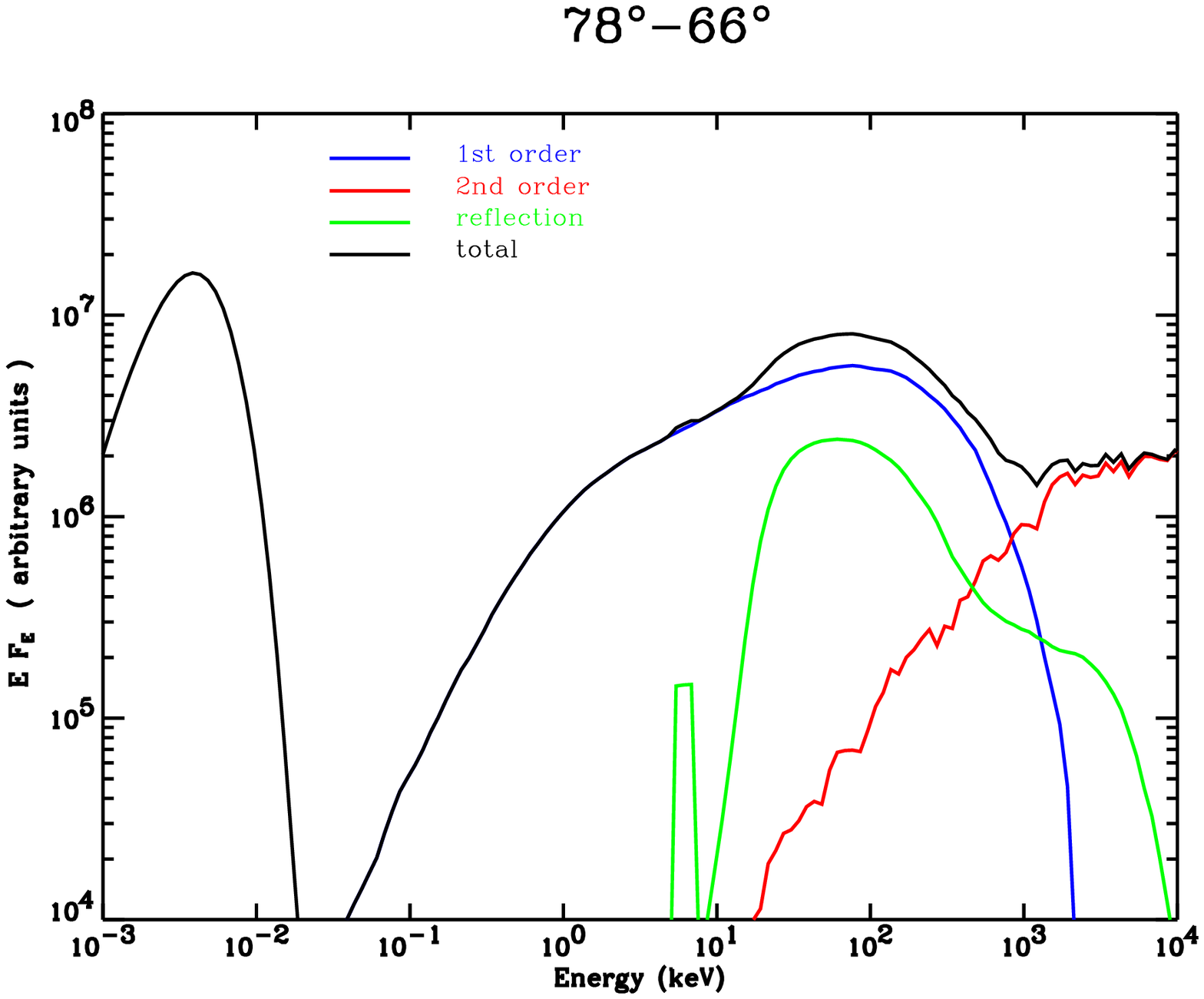,width=7cm}}
\centerline{\psfig{file=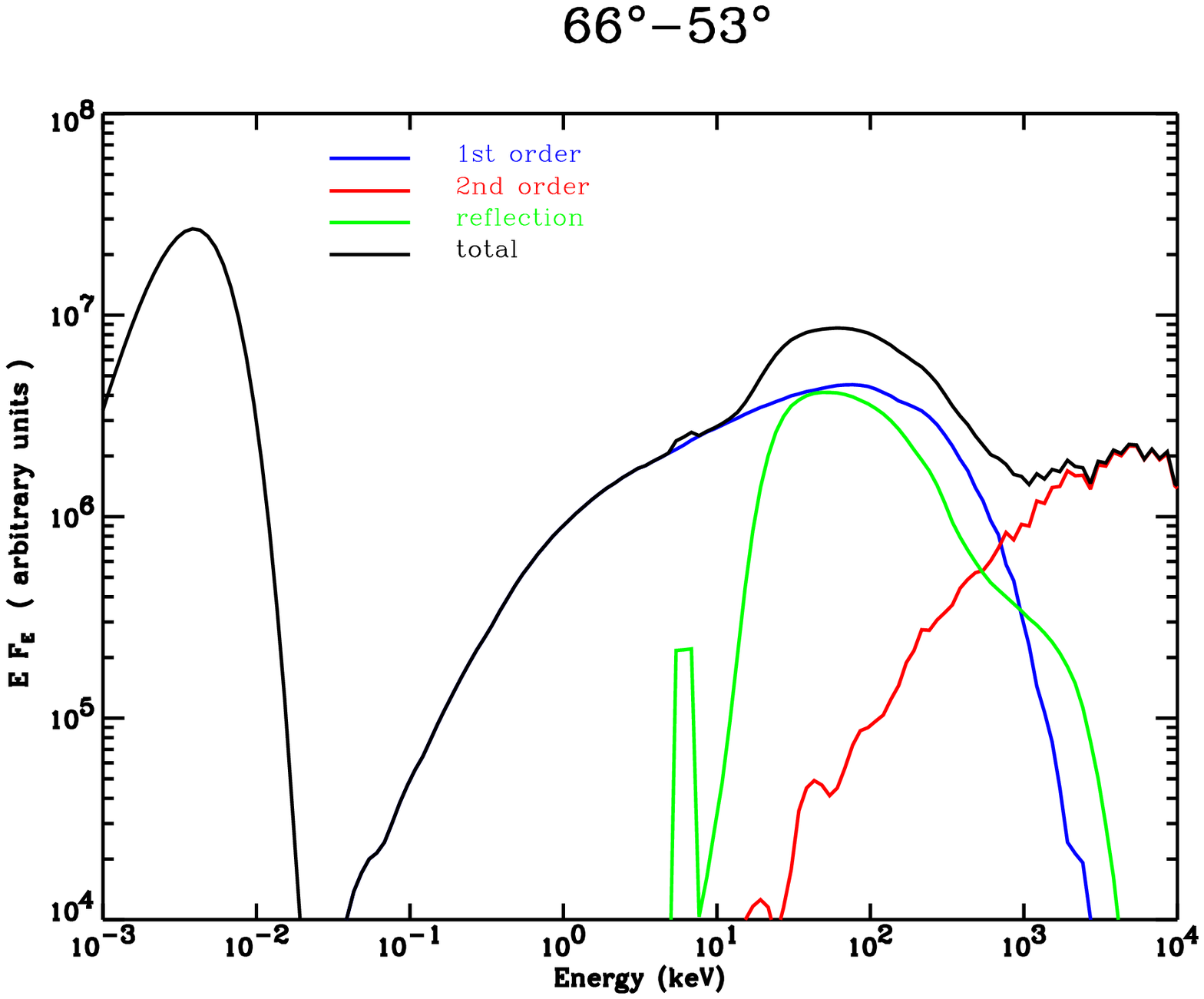,width=7cm}\psfig{file=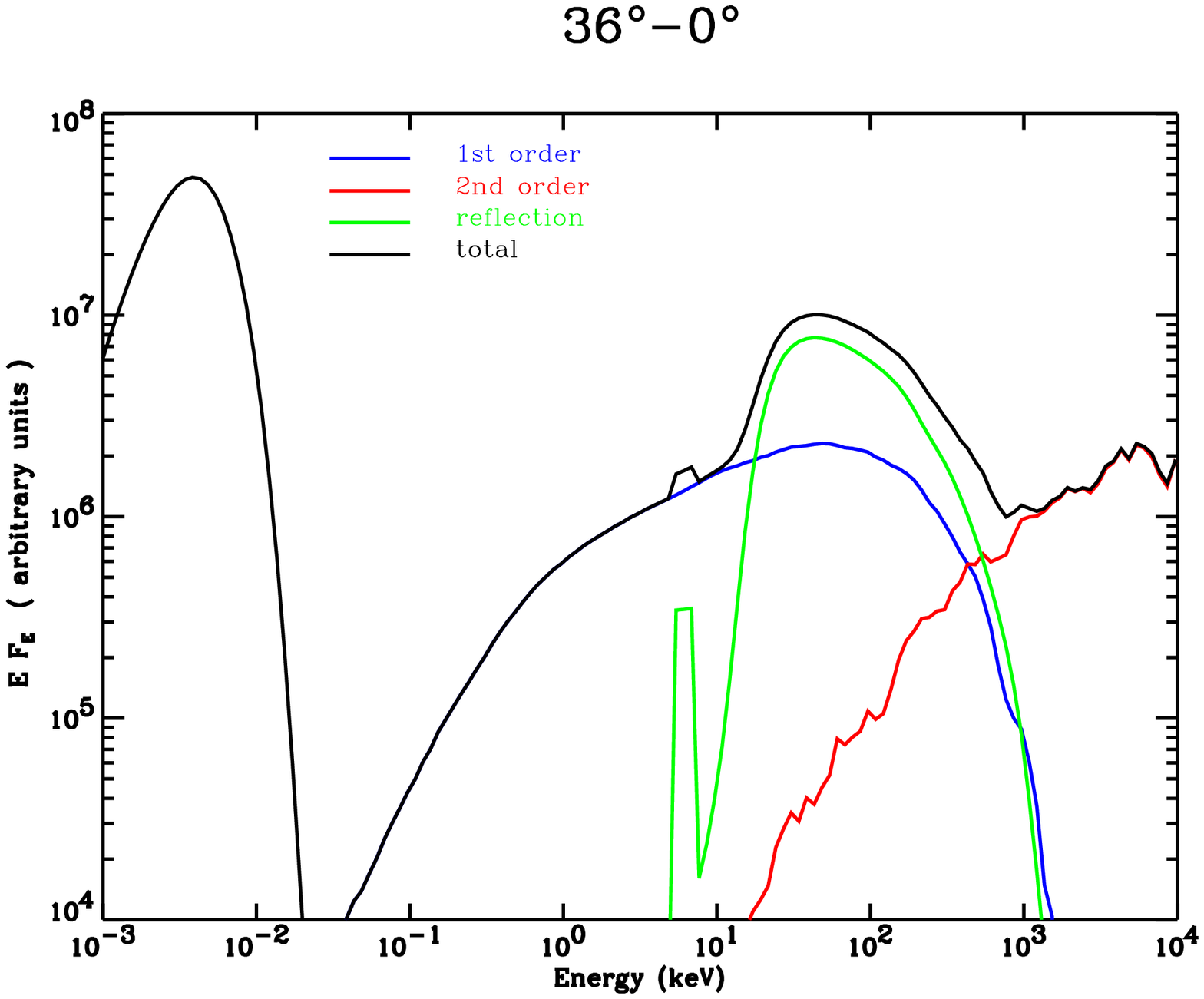,width=7cm}}
\centerline{\caption{FIGURE 5. $s=2.1$}}
\end{figure}

$kTr$ as pratically no influence on K provided it is lower than 0.1 keV.
$K$ is very sensible to the mean lorentz factor of the electron distribution. Fig. 4 shows $K$ versus $<\gamma>$ for $\chi=1$, $cos\theta=.83$, $kTr=50 eV$ and various values of $\gamma_{min}$, $\gamma_{0}$, $\beta$ and $s$.
Clearly, $K$ is minimum for a slab coronna geometry ($\chi=0$ and $cos\theta=0$) and large values of $<\gamma>$.
We will now focus on an extreme case of corona geometry to study a configuration which provides the lowest  achievable $K$ values. We fixed $\gamma_{0}=200$ (its maximal observational limit compatible with $kTr=1 eV$). 
Fig. 2 shows the evolution of $K$ as a function of $\gamma_{min}$ for $s=2.1$ (upper curve) and $s=3$ (lower curve) ($\chi=2.\,10^{-3}$, $cos\theta=0$, $\gamma_{0}=200$, $\beta=3.$). Note that in this limit of low $\chi$, $K\sim\tauh$. In the lower region the 2nd CSO is negligible for $s<3.$; in the middle region it is negligible at least for $s=2$. In the upper and darker region the 2nd CSO is never negligible for $2<s$. These regions have been determined qualitatively by simulations of a slab geometry with the linear code. The green region can not be reached even for the extreme value of $\gamma_{min}=10$. However, the curve $s=2.1$ crosses the middle region. We thus see that the model
can only reproduce the spectra with the hardest observed slopes.
\begin{figure}[h]
\centerline{\psfig{file=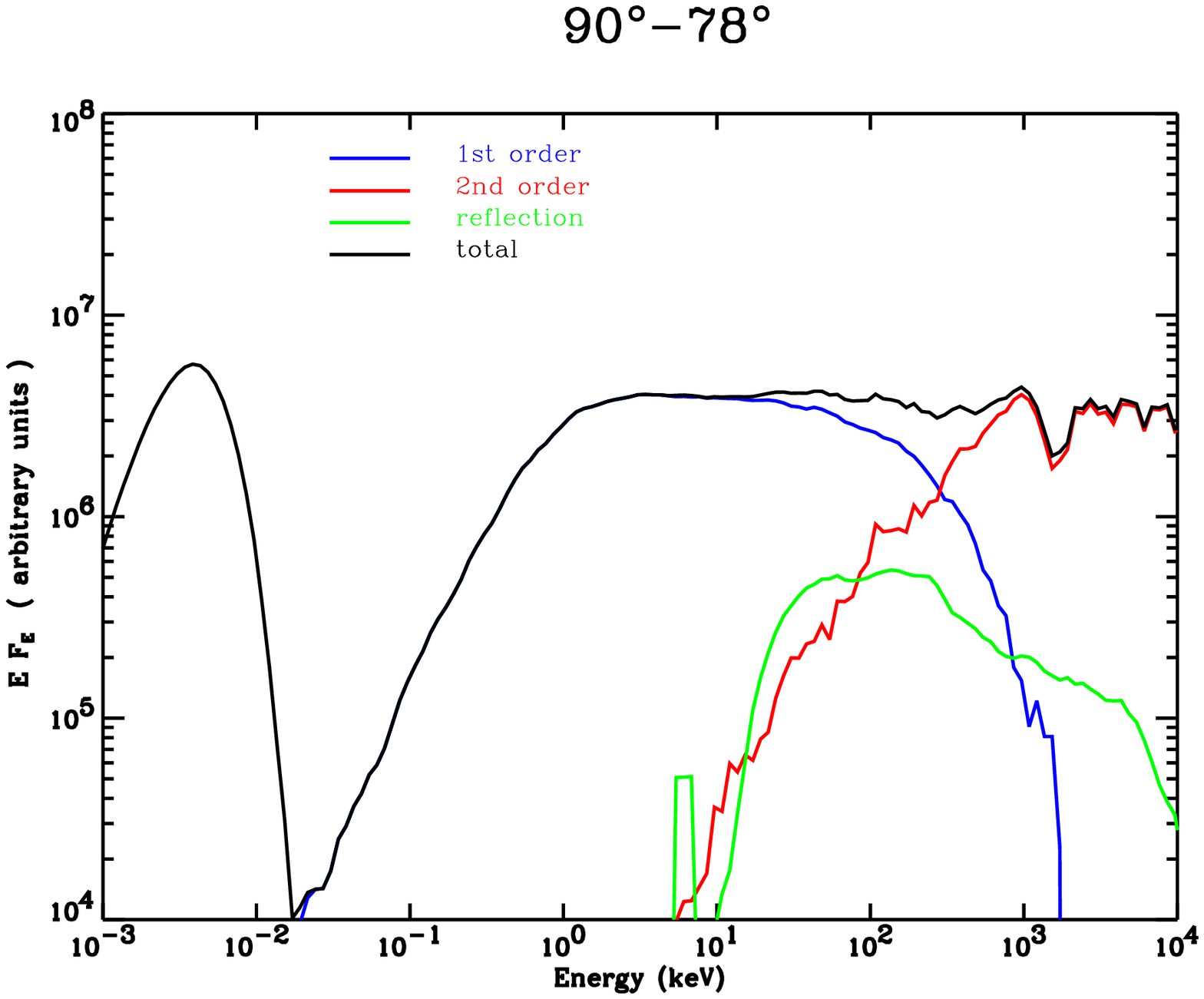,width=7cm}\psfig{file=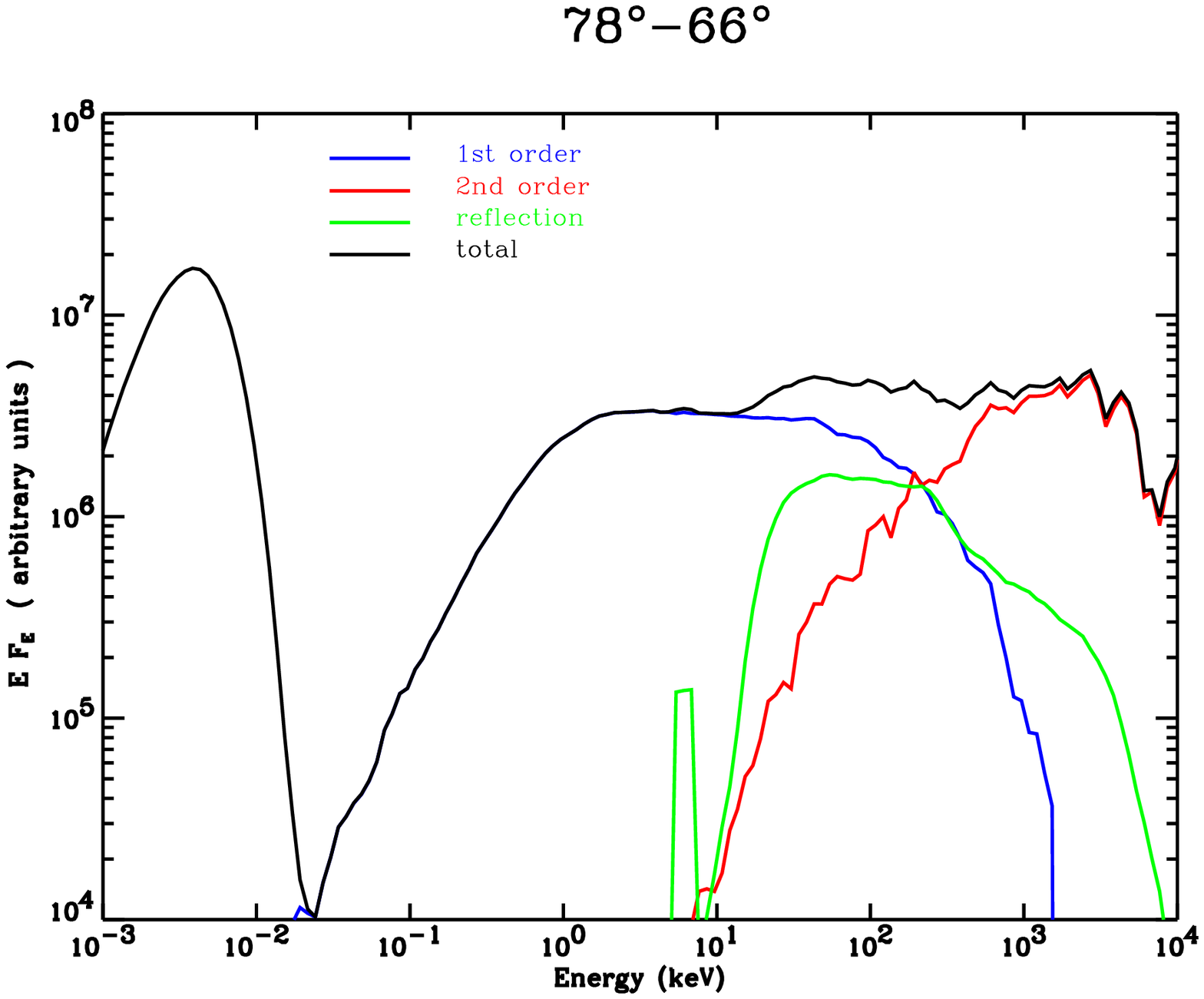,width=7cm}}
\centerline{\psfig{file=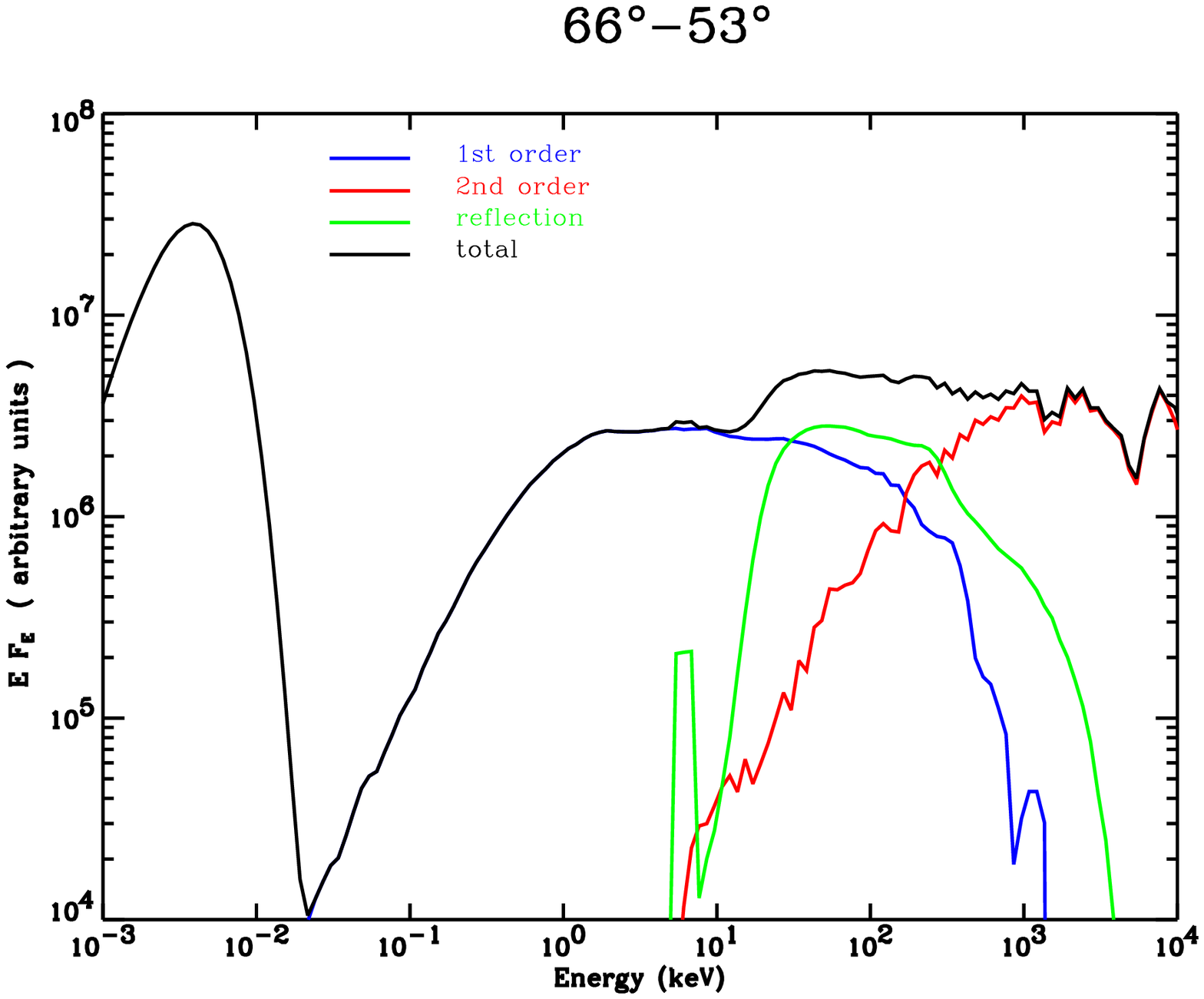,width=7cm}\psfig{file=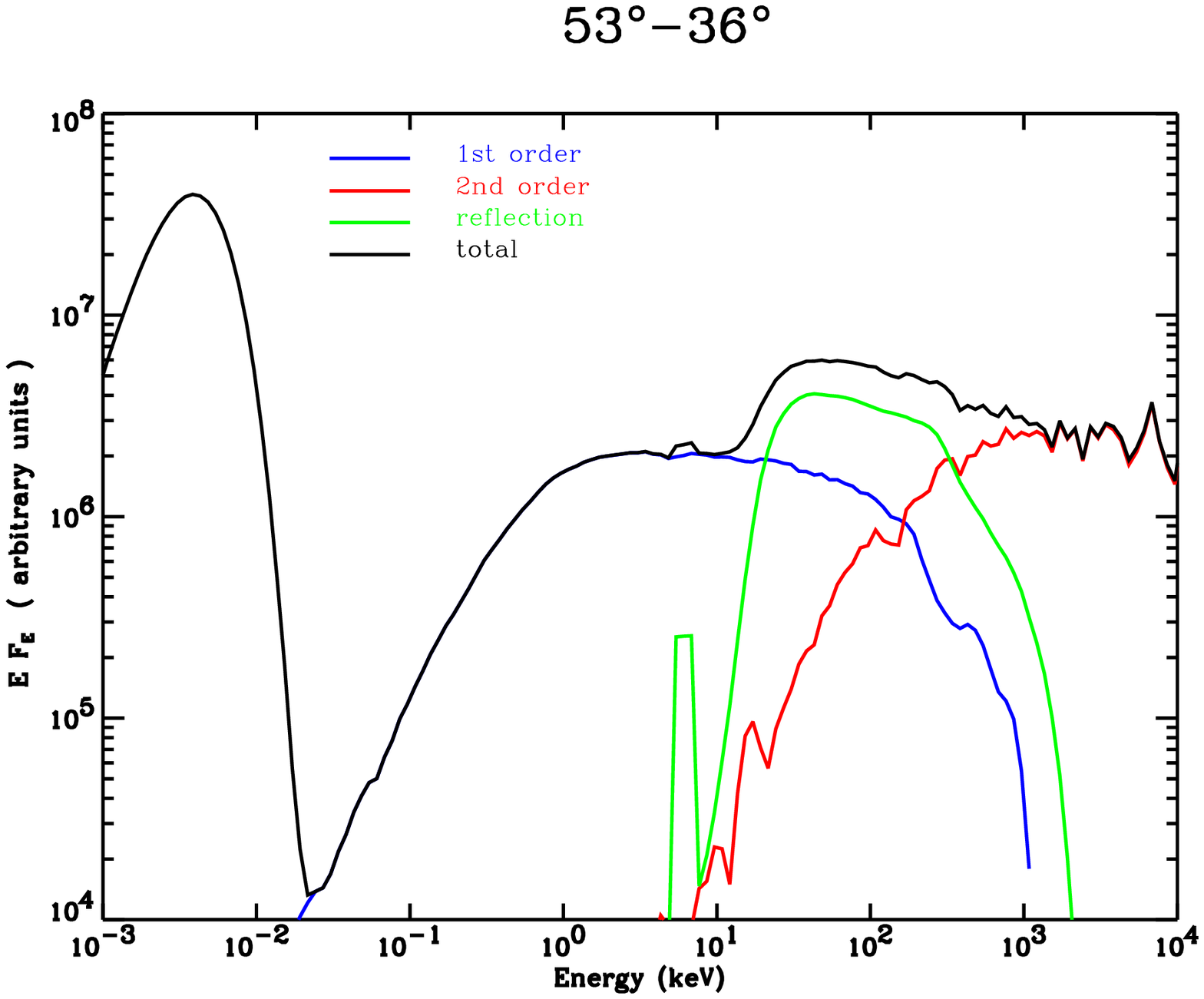,width=7cm}}
\centerline{\caption{FIGURE 6. $s=3$}}

\end{figure}
 
Fig 5 shows the spectra obtained
with $s=2.1$, $\gamma_{min}=10$ ($\tauh\sim4\,10^{-4}$) for different inclination angles. The 2nd CSO is well below the first order for $E <1$ MeV. It becomes dominant above 1 MeV, but at these energies the flux can be considerably lowered by photon-photon absorption. 
Fig. 6 shows the spectra obtained
with $s=3$, $\gamma_{min}=10$ ($\tauh\sim8.\,10^{-4}$) for different inclination angles. The 2nd CSO is important for all viewing angles.
The spectrum resulting in the sum of the two CSOs does not show any cut-off around 200 keV.     
These spectra show that the previous results concerning the angular dependence of the spectra (HP97) are not sensibly affected by the source extension.
The reflection component appears a bit lower, relaxing somewhat the constraints on the inclination angle (Malzac et al., 1998). However the spectra are still reflection dominated at small inclinations.

\bsk
\ni  
4. CONCLUSION
\ni
\ssk
\ni

Our self consistent simulations showed that low optical depths could be reached only if the source is very close to the cold disc (corona-like geometry). Moreover, even in this case, the 2nd CSO is negligible only for the hardest observed spectral slopes $\alpha\sim1.6$. Lower optical depths could be obtained if consequent dissipation occured in the 
disc.
However, the source extension does not suppress the unobserved dominent reflection component at small inclinations. 

\ssk
\baselineskip = 12pt

{\references \ni REFERENCES
\ssk
      Henri, G. \& Petrucci, P. O., 1997, A\&A, 326, 87, (HP97)

%

      Podzniakov, I. A., Sobol, I. M., \& Sunyaev, R. A., 1983, Ap Space Phys. Review, 2, 189

      Stern, B.E., Poutanen, J., Svenson, R., Sikora, M., Begelman, M.C., 1995, ApJ, 449, L13

      Malzac, J., Jourdain, E., Petrucci, P.O., Henri, G., 1998, A\&A, 336, 807 
}

\end{document}